\documentclass[prb,twocolumn,showpacs,floatfix]{revtex4}
\usepackage{graphicx}
\usepackage{dcolumn}
\usepackage{bm}

\begin{document}

\preprint{}

\title{Quantum states and linear response in dc and electromagnetic fields for charge current
       and spin polarization of electrons at Bi/Si interface with giant spin-orbit coupling}

\author{D.V. Khomitsky\footnote{E-mail: khomitsky@phys.unn.ru}}

\affiliation{Department of Physics, University of Nizhny Novgorod,
              23 Gagarin Avenue, 603950 Nizhny Novgorod, Russian Federation}

\date{\today}

\begin{abstract}

An expansion of the nearly free-electron model constructed by Frantzeskakis, Pons and Grioni
[Phys. Rev. B {\bf 82}, 085440 (2010)] describing quantum states at Bi/Si(111) interface
with giant spin-orbit coupling is developed and applied for the band structure and
spin polarization calculation, as well as for the linear response analysis for charge current
and induced spin caused by dc field and by electromagnetic radiation. It is found that
the large spin-orbit coupling in this system may allow resolving the spin-dependent properties
even at room temperature and at realistic collision rate. The geometry of the atomic lattice combined
with spin-orbit coupling leads to an anisotropic response both for current and spin components related
to the orientation of the external field. The in-plane dc electric field produces only the in-plane
components of spin in the sample while both the in-plane and out-of-plane spin components can be
excited by normally propagating electromagnetic wave with different polarizations.

\end{abstract}


\maketitle

\section{Introduction}

It is well-known that the knowledge of a material with high values of spin-orbit (SO) coupling
parameters is a goal for many theoretical, experimental, and device research groups in condensed
matter physics and spintronics due to its fascinating spin-related properties and
possible applications in the information processing and storage. Among the candidates which
attract considerable attention throughout the last decade there are Bi/Si(111) surface alloys
which band structure has been experimentally studied for several years and recently
modeled in a paper \cite{FPG} by Frantzeskakis, Pons and Grioni. This material, in line
with the other examples of "metal-on-semiconductor" systems with large SO coupling, has been a subject of
many both experimental and theoretical papers since it seems very promising to make use of a material combining
the large SO splitting of Bi and conventional semiconductor technology of Si which is one of the main goals
of spintronics.\cite{Awschalom,Zutic,Wu} Here we shall mention only some of the numerous results of research on
the Bi-covered Si surface properties with various crystal orientation of Si substrate. In particular, the scanning
tunneling microscopy has been used to determine the surface structure of Bi/Si some 18 years ago, \cite{Shioda1993}
and the analysis of the atomic geometry and electronic structure continued further \cite{Miwa2002,Miwa2003} focusing
throughout the recent years mainly on the atomic surface geometry and spin-resolved band structure reconstruction
where the methods of angle-resolved photoemission spectroscopy have been applied
\cite{FPG,Hirahara2006,Hirahara2007,Hirahara2007b,Grioni2008,Dil2008,Gierz2009,Bian2009,Sakamoto2009}.
Other methods included the low-energy electron diffraction and atomic force microscopy
\cite{Jnawali2006,Jnawali2007}, and besides pure Si, the Si-Ge superlattices have been used
as a substrate for Bi coverage, \cite{Miwa2005} and later the lateral Ge-Si nanostructure
prepared on the Si/Bi surface have been studied by the scanning tunneling
microscopy.\cite{Myslivecek2010} Also, for the Bi/Si system there were studies of energetic stability and
equilibrium geometry \cite{Miwa2006} and the possibility of designing iron silicide wires along Bi
nanolines on the hydrogenated Si surface,\cite{Miwa2008}, and of the thermal response upon the femtosecond
laser excitation. \cite{Hanisch2008} It is well-known that Bi is a material with very
big SO splitting, and thus it attracts a steady interest in its potential applications in
spintronics where various schemes of combining it with semiconductors are suggested, one of the most
recent being an investigation of BiTeI bulk material where the SO splitting reaches a very high value
of $0.4$ eV.\cite{Nagaosa2011}

From the list of papers mentioned above it is evident that the geometric properties of atom arrangement
and the resulting band structure have already been studied for Bi/Si systems by many experimental and theoretical
groups. However, much lower attention have been given so far to the prediction and observation of different effects
caused by the electron system response to an external excitation, including such basic properties as the charge current
and spin polarization in the dc field which are often are considered as the starting point of the response calculations,
especially for systems with important role of SO coupling.\cite{Edelstein1990,Aronov1991,Kleinert2005,Raichev2007}
Besides the response to the dc electric field, the optical properties of SO-split band spectrum always attracted
significant attention starting from the conventional semiconductor structures with big SO
coupling.\cite{Bhat2005,Sherman2005,Tarasenko2005,Ganichev2007,Berakdar2010}
In our previous papers we have observed an important role of SO coupling in
conventional InGaAs-based semiconductor superlattices on the energy band formation \cite{DK2006}
which directly affected both charge and spin response for the excitation by the electromagnetic
radiation \cite{sooptic,PSK} and by the dc electric field. \cite{spinelec} It is known that the spin polarization
configurations in semiconductors may have a rather long relaxation time \cite{Wu,Pershin2005,Stano2006}
which makes them as important as the conventional charge current setups for their applications in
the nanoelectronics and spintronics.

While there is no doubt that the electron properties in Bi-covered Si interface
differ from the ones in conventional semiconductor structures with strong SO coupling,
the questions regarding their SO-dependent response to the external dc and electromagnetic fields
remain to be very important since we are still in the beginning of our way towards
understanding and utilizing such novel materials with giant SO coupling as Bi/Si.

The goal of the present paper is to apply a modified and expanded version of simple but adequate nearly free electron
model\cite{FPG} for the band structure of electrons in Bi covering the Si(111) surface
which allows us to calculate various physical characteristics of this material in the linear response regime,
including the charge current and spin polarization caused by dc electric field with
different orientations, and also to obtain the response of non-equilibrium spin
polarization excited by the electromagnetic field with various polarizations.
Indeed, such physical quantities can be among the first ones measured in the nearest
experiments on Bi/Si, and hence it is of interest to calculate them beforehand both qualitatively and,
when possible, quantitatively. Since we do not know exactly for the present moment
many material parameters of the electron system in Bi-covered Si interface, including
such parameters as the electron surface concentration and mobility, the dielectric
tensor, the relaxation rates for charge and spin, etc, we sometimes cannot yet
calculate the effects in the absolute measurable units and use the standard label "arbitrary
units" instead. Still, we believe that the comparison of the output results for the
same physical parameter calculated at different conditions always has a value since
it allows to predict the relative significance of them when the conditions are varied.
We find both common and distinct features of the charge and spin
system response in Bi/Si compared to well-known GaAs or InGaAs semiconductor
structures. Thus, we believe that our findings can be a good starting point for further
theoretical and experimental studies of charge and spin response in Bi/Si system at
various external fields.

The paper is organized as follows: in Sec.II we introduce the expanded version of the nearly free electron
Hamiltonian for Bi/Si and discuss the band structure and spin polarization in the energy bands,
in Sec.III we solve the kinetic equation and calculate the charge current and spin polarization in
the dc electric field, in Sec.IV we study the excited spin polarization in the framework of
the linear response theory for different polarizations and frequencies of the incident electromagnetic wave,
and we give our conclusions in Sec.V.

\section{Hamiltonian and quantum states}

\begin{figure}
  \centering
  \includegraphics[width=85mm]{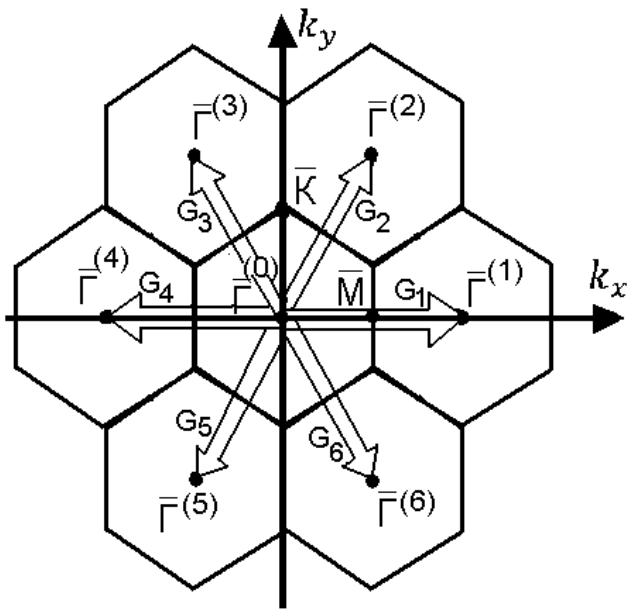}
  \caption{Surface Brillouin zones (SBZ) of Bi/Si with a hexagonal shape shown by solid lines and
           the reciprocal lattice (hollow) vectors ${\bf G}_1, \ldots, {\bf G}_6$ connecting the
           equivalent Gamma-points  ${{\overline \Gamma}}^{(0)},\ldots , {{\overline \Gamma}}^{(6)}$
           of the nearest neighbor approximation for the nearly free electron model.
           The spin-split Rashba paraboloids are centered in each of the Gamma-points.}
  \label{fsbz}
\end{figure}

Recent proposal of several theoretical models for band structure calculation of Bi/Si
electron surface states \cite{FPG} provided a variety of choices for studies of the
corresponding Hamiltonian and quantum states. Here we shall start with the simplest
nearly free electron (NFE) model which was proposed initially for the description of
the band structure mainly in the vicinity of the $\overline{M}$ point at the surface
Brillouin zone (SBZ) of Bi/Si having a hexagonal shape shown by dashed contour in Fig.\ref{fsbz}.
We shall only briefly describe it here since the detailed derivation and discussion is
available in the original paper.\cite{FPG}
The choice of the reciprocal lattice vectors initially has been restricted to three
vectors ${\bf G}_1$, ${\bf G}_2$ and ${\bf G}_3$ connecting four equivalent
Gamma-points ${\overline \Gamma}^{(0)}$, ${\overline \Gamma}^{(1)}$, ${\overline \Gamma}^{(2)}$ and
${\overline \Gamma}^{(3)}$. In the framework of the NFE approach for each Gamma-point the standard
$2 \times 2$ Rashba Hamiltonian of a free electron in the $\hat{\sigma}_z$ basis has been written with
the center of the quasimomentum at the corresponding ${\overline \Gamma}^{(n)}$ point. As a result,
an $8 \times 8$ matrix is derived giving the energy bands and two-component eigenvectors (the Rashba spinors)
describing the spin polarization in the reciprocal space.

We are going to use an expanded version of this model by including the remaining
reciprocal vectors ${\bf G}_4$, ${\bf G}_5$ and ${\bf G}_6$ into our basis of nearest
neighbor sites connecting the center Gamma-point  ${\overline \Gamma}^{(0)}$ with all
surrounding points ${\overline \Gamma}^{(1)},\ldots , {\overline \Gamma}^{(6)}$, as it is shown by hollow
vectors  ${\bf G}_1, \ldots, {\bf G}_6$ in Fig.\ref{fsbz} where several hexagonal SBZ-s are shown by solid contours,
thus creating a $14\times 14$ Hamiltonian matrix. We assume the previously determined\cite{FPG} values of geometrical
parameters $\overline{\Gamma M}=0.54$ ${\AA}^{-1}$ and $\overline{\Gamma K}=0.62$ ${\AA}^{-1}$. Such an expansion
allows us to treat a much wider area of the SBZ compared to the region near the $\overline{M}$-point \cite{FPG}
and to keep the symmetry of the non-trivial hexagonal Bi/Si(111) trimer structure with one monolayer
of Bi atoms.\cite{FPG,Miwa2002,Miwa2003,Gierz2009}

Our Hamiltonian may thus be described via its matrix elements in the following form:

\begin{equation}
H_{nn'}=H_R({\bf k}+{\bf G}_n) \delta_{nn'} + V_{nn'},
\label{ham}
\end{equation}

and the electron spinor wavefunction is constructed as

\begin{equation}
\Psi_{{\bf k}}({\bf r})=\sum_n a_{n{\bf k}} \psi_{n{\bf k}}({\bf r})
\label{wf}
\end{equation}

where the conventional form of Rashba Hamiltonian is used,

\begin{equation}
H_R({\bf k})=
\left(
\begin{array}{cc}
\hbar^2 k^2/2m & \alpha_R(k_y+i k_x) \\
\alpha_R (k_y-i k_x) & \hbar^2 k^2/2m
\end{array}
\right),
\label{hr}
\end{equation}

and the matrix elements of the periodic potential coupling the free electron states are

\begin{equation}
V_{nn'}=
\langle \psi_n \mid \sum_m V_0 \exp (i {\bf G}_m {\bf r}) \mid  \psi_{n'} \rangle.
\label{vnm}
\end{equation}

The basis functions $\psi_{n{\bf k}}({\bf r})$ in Eq.(\ref{wf}) are the well-known Rashba spinors
$\psi_{n{\bf k}} = \psi_{\bf{k}+\bf{G}_n}$ where

\begin{equation}
\psi_{{\bf k}}=\frac{e^{i {\bf kr}}}{\sqrt{2}}
\left(
\begin{array}{c}
1 \\
\pm e^{i {\rm Arg} (k_y-i k_x)}
\end{array}
\right),
\label{rf}
\end{equation}

and the $(\pm)$ sign corresponds to two eigenvalues for Rashba energy spectrum
$E(k)=\hbar^2 k^2/2m \pm \alpha_R k$. We continue to adopt here the known values of
material parameters for Eqs (\ref{ham})-(\ref{vnm}) and constants used during the initial
construction\cite{FPG} of the NFE model for Bi/Si, namely, we put
$m=0.8 m_0$, $\alpha_R=1.1$ $\rm{eV \cdot \AA}$ and $V_0=0.3$ eV.

\begin{figure}
  \centering
  \includegraphics[width=85mm]{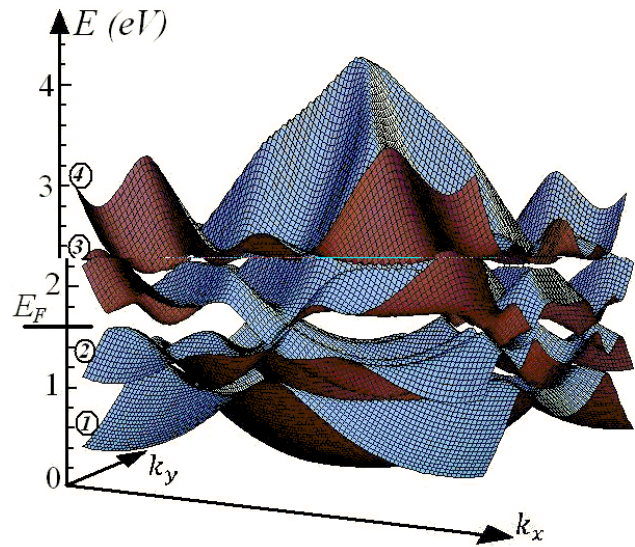}
  \caption{(Color online) Energy band structure of the electrons on Bi-covered Si(111) surface for
           the four lowest bands labeled from 1 to 4. The Fermi level is located at $E_F=1.6$ eV
           between the band No.2 and band No.3 where a small global energy gap of around $0.1$ eV is formed.
           The cross-sections of our 3D band plot shown here accurately repeat the 2D plots for the energy
           dispersion lines along various directions in the SBZ which were studied earlier in the framework
           of the NFE model.\cite{FPG}}
  \label{fbands}
\end{figure}

After diagonalization of the Hamiltonian (\ref{ham}) we arrive to the energy band
spectrum $E=E_s(k_x,k_y)$ where $s=1,2,\ldots$ labels the energy bands of the electrons
in the Bi/Si(111) system. The three-dimensional plot of the energy band structure is
presented in Fig.\ref{fbands} for the four lowest bands labeled from 1 to 4. These
lowest bands seem to be of the primary importance for the electron response analysis
since the Fermi level is reportedly located\cite{FPG} in the middle of them at $E_F=1.6$ eV,
i.e. between the band No.2 and band No.3, as it can be seen in Fig.\ref{fbands}.
One of the most important features of the spectrum in Fig.\ref{fbands} stemming from the lattice geometry
is the hexagonal symmetry of the energy bands in ${\bf k}$-space which implies, among
other things, the absence of the $k_x \leftrightarrow k_y$ symmetry, leading to rich
properties of the spin response phenomena as we shall see below.
It should be mentioned that the cross-sections of our 3D band plot shown here accurately repeat
the 2D plots for the energy dispersion lines along various directions in the SBZ which were studied
earlier in the framework of the NFE model.\cite{FPG} We would like to add here just some new quantitative data:
the cross-sections of the energy band surfaces along the ${\overline \Gamma}- {\overline M}$ direction reported
previously\cite{FPG} might created an impression of a large energy gap in the whole spectrum between
bands 2 and 3 while the complete 3D presentation of these bands in Fig.\ref{fbands} indicates that the global
structure of the energy bands in the whole 2D SBZ leaves this gap opened but with a much smaller
width of around $0.1$ eV. Of course, the precise values of energy gaps may vary from model to model
and can be specified more precisely during the future experimental and theoretical analysis.

Another important characteristic of quantum states in a system with significant SO
coupling is the spin polarization of the eigenstates $\psi_{{\bf k}}$ in the Brillouin zone
which can be defined as the vector field in the reciprocal space with the components ($m=x,y,z$)

\begin{equation}
S_{m}({\bf k})=
\langle \psi_{{\bf k}} \mid {\hat \sigma}_m \mid  \psi_{{\bf k}} \rangle.
\label{sm}
\end{equation}

\begin{figure}
  \centering
  \includegraphics[width=80mm]{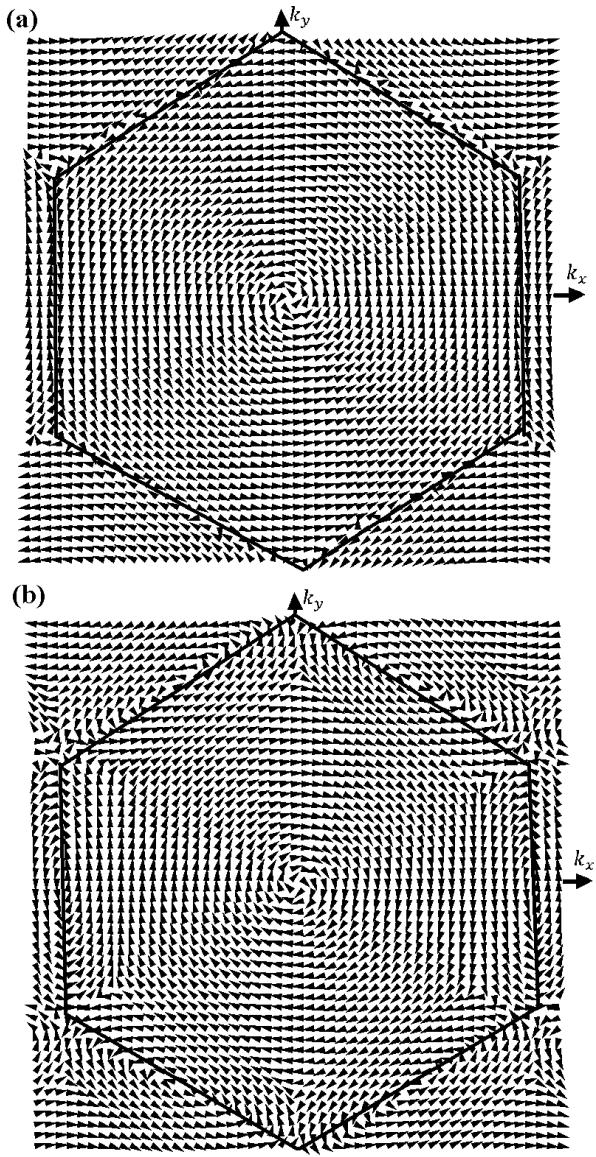}
  \caption{Two-dimensional spin polarization distributions $(S_x({\bf k}),S_y({\bf k}))$ for the lowest
           energy bands No.1 (a) and No.2 (b) from the band spectrum shown in Fig.\ref{fbands}, with
           the hexagonal SBZ marked by a solid contour. The initial Rashba counter clock-wise and clock-wise
           patterns are present in a wide area surrounding the SBZ center, but more complicated vector field
           structure near the SBZ edge.}
  \label{fs12}
\end{figure}

\begin{figure}
  \centering
  \includegraphics[width=80mm]{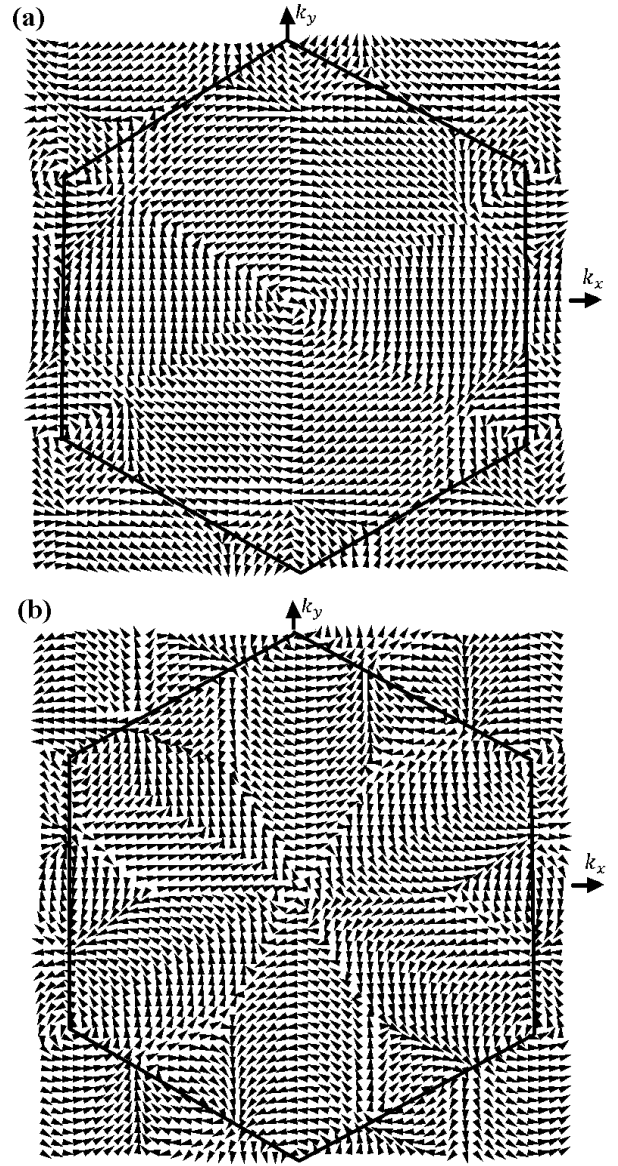}
  \caption{Spin polarization distributions $(S_x({\bf k}),S_y({\bf k}))$ for the higher
           energy bands No.3 (a) and No.4 (b) from the band spectrum shown in Fig.\ref{fbands}, with the hexagonal
           SBZ marked by a solid contour. The spin polarization in band No.3 (a) and especially in band No.4 (b)
           demonstrates more new properties compared to the free Rashba states, including
           both the shape of the spin vector field which captures now more features of the hexagonal
           geometry of the SBZ, and the arising of new local vortices at various points of
           symmetry of the SBZ, mainly near its corners.}
  \label{fs34}
\end{figure}

As usual for the Hamiltonian with a pure Rashba SO coupling term, the out-of-plane
component $S_z$ of the spin field vanishes. The remaining components form a 2D spin
polarization distribution in the SBZ which forms a specific vector field picture for each
of the energy bands. In Fig.\ref{fs12} and Fig.\ref{fs34} we show the 2D spin
polarization distributions $(S_x({\bf k}),S_y({\bf k}))$ for the lowest energy bands
No.1 (Fig.\ref{fs12}a), No.2 (Fig.\ref{fs12}b), No.3 (Fig.\ref{fs34}a) and No.4 (Fig.\ref{fs34}b)
from the band spectrum shown in Fig.\ref{fbands}, with the hexagonal SBZ marked by a solid contour.
As for the spins in two lowest subbands shown in Fig.\ref{fs12}, one can see here that the initial Rashba
counter clock-wise and clock-wise patterns of spin directions are present in a rather wide area surrounding
the SBZ center, but more complicated vector field structure arises near the SBZ edge.
The spin polarization in higher band No.3 and especially in band No.4 shown in Fig.\ref{fs34}
demonstrates more new properties compared to the free Rashba states, including both
the shape of the spin vector field which captures now more features of the hexagonal
geometry of the SBZ, and the arising of new local vortices at various points of
symmetry of the SBZ, mainly near its corners.

It can be concluded from the analysis of the spin polarization in the energy bands of Bi/Si system
that certain properties of the initial basis of Rashba states remain visible. However, the change of the space
symmetry to the hexagonal type without the element of axial symmetry and without the $x \leftrightarrow y$ symmetry
may probably lead to both common and distinct features in the current and spin response to the application
of various external fields compared to the well-known properties of 2DEG with Rashba
SO coupling. This assumption will be confirmed and illustrated below.

\section{Charge current and spin polarization response for dc field}

It is known that the response of a two-dimensional electron gas with SO coupling to a
constant electric field may be accompanied not only by the charge current but also by
the spin polarization. \cite{Kleinert2005,Raichev2007,spinelec} The most significant
properties of such response for a pure Rashba SO term (\ref{hr}) in the Hamiltonian is
the arising of the in-plane transverse polarization, i.e. the $S_{y(x)}$ spin component
when the electric field is applied along the $x(y)$ direction while the out-of-plane $S_z$
component is absent in case of the accurately included relaxation processes which is
sometimes related also to the absence of the spin Hall effect for a ${\bf k}$-linear Rashba
model in the presence of the disorder.\cite{Inoue2004} So, it is natural to start the
analysis of the electron system response with the calculation of the charge current
and field-induced spin.

We shall start with the calculation of the non-equilibrium electric field-affected distribution
function ${\tilde f}_m({\bf k},E_i)$ in the m-th energy band. If the system is subjected to a constant and
uniform electric field $E_i$ parallel to the i-th axis, then in the collision
frequency approximation the kinetic equation for ${\tilde f}_m({\bf k},E_i)$ can be written
as \cite{spinelec}

\begin{equation}
eE_i \frac{\partial {\tilde f}_m({\bf k},E_i)}{\partial k_i}=-\nu [{\tilde f}_m({\bf k},E_i)-f_m({\bf k})],
\label{kineq}
\end{equation}

where $\nu$ is the collision rate and $f_m({\bf k})=1/(1+\exp[(E_m({\bf k})-\mu)/k_B T])$
is the Fermi equilibrium distribution function in the $m$-th band. Since the Bi/Si energy spectrum
is characterized by a very large SO splitting and the band widths of the order of 1 eV, it may be
a promising candidate for spin-dependent phenomena visible at room temperature. Thus, in the following
we shall assume that $T=293$ K and consider a value of $\nu=10^{12}$ $s^{-1}$. As we have said before,
the estimate for the collision rate as well for many other material parameters for the Bi/Si system is
presently based on the assumptions rather than of the solid experimental facts since
we are still in the beginning of the investigation for this new material. Still, we
believe that our qualitative and sometimes quantitative results can be useful for
predicting some novel properties of the electron and spin system response.

The mean charge current density $j_i(E_i)$ measured for 2D system in units of current divided
by the unit of transverse system size and the mean spin polarization $S_j$, $j=x,y,z$,
can be found as \cite{spinelec}

\begin{equation}
j_i(E_i)=e n \sum_{m,{\bf k}} \langle \psi_{m{\bf k}} \mid v_i \mid \psi_{m{\bf k}}\rangle
                            {\tilde f}_m({\bf k},E_i),
\label{ji}
\end{equation}

\begin{equation}
S_j(E_i)=\sum_{m,{\bf k}} \langle \psi_{m{\bf k}} \mid \sigma_j \mid \psi_{m{\bf k}} \rangle
                               {\tilde f}_m({\bf k},E_i),
\label{sn}
\end{equation}

where $n$ is the surface concentration of the electrons on the Bi-covered Si surface, $\sigma_i$
($i=x,y,z$) are the Pauli matrices, and the velocity operator $v_i=\partial {H} / \partial k_i$
includes the SO part proportional to the Rashba parameter $\alpha_R$ and acts on the spinor wavefunctions (\ref{wf})
via the matrices

\begin{equation}
v_x=
\left(
\begin{array}{cc}
-i\hbar\nabla_x/m & i\alpha/\hbar \\
-i\alpha/\hbar & -i\hbar\nabla_x/m
\end{array}
\right),
\label{vx}
\end{equation}

\begin{equation}
v_y=
\left(
\begin{array}{cc}
-i\hbar\nabla_y/m & \alpha/\hbar \\
\alpha/\hbar & -i\hbar\nabla_y/m
\end{array}
\right).
\label{vy}
\end{equation}

In order to take a closer look for onto the expectations of the charge current density, we shall
calculate $j_x(E_x)$ and $j_y(E_y)$ as well as $S_i(E_x)$ and $S_i(E_y)$ ($i=x,y,z$)
by applying Eq.(\ref{ji}) for the electron surface concentration $n=10^{14}$ ${\rm cm}^{-2}$ which may be
reasonable since the surface density of atoms on Bi-covered Si(111) surface is estimated
on the order of $10^{15}$ ${\rm cm}^{-2}$ according to the experimental observations.\cite{Hirahara2007,Hirahara2007b}

\begin{figure}
  \centering
  \includegraphics[width=75mm]{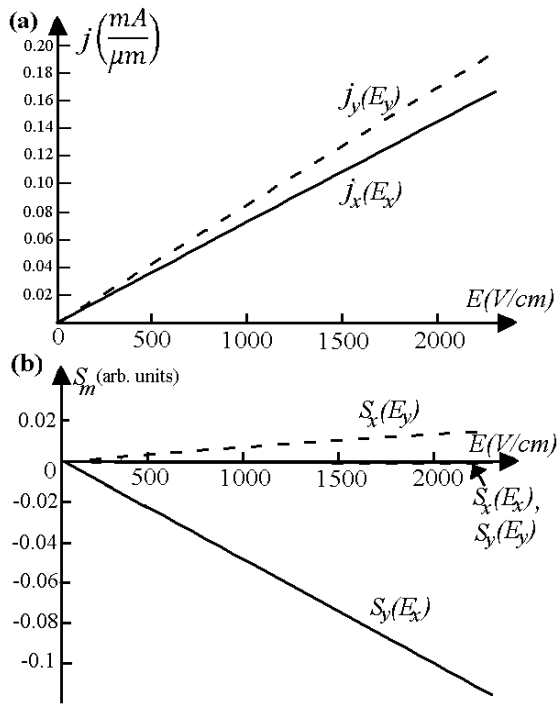}
  \caption{(a) Charge current density (\ref{ji}) and (b) spin polarization (\ref{sn}) induced by
           the dc electric field applied along $x$ (solid curves) and $y$ (dashed curves).
           A conventional linear dependence of the charge current on the applied electric field throughout
           the whole range of fields and almost linear dependence for the significant non-zero
           induced spin components $S_y(E_x)$ and $S_x(E_y)$ are visible. Other in-plane components $S_x(E_x)$
           and $S_y(E_y)$ marked by arrow also present but their magnitude is much lower, and the out-of plane
           $S_z$ component is negligibly small.}
  \label{fjs}
\end{figure}

The results for the charge current density (\ref{ji}) and spin polarization (\ref{sn}) are
shown in Fig.\ref{fjs}(a) and Fig.\ref{fjs}(b) respectively. One may see a conventional
linear dependence of the charge current on the applied electric field throughout the whole range of
fields up to $2$ kV/cm, and almost linear dependence for the significant non-zero induced spin components
$S_y(E_x)$ and $S_x(E_y)$. Other in-plane components $S_x(E_x)$ and $S_y(E_y)$ marked by arrow
are also present in Fig.\ref{fjs} but their magnitude is much lower compared to $S_y(E_x)$ and
$S_x(E_y)$, and the out-of plane $S_z$ component is negligibly small.
It is evident that the lattice asymmetry with respect to the $x \leftrightarrow y$
interchange transform has lead to a slight but distinct asymmetry in the current amplitude of around $12 \%$,
and the dominating $S_y(E_x)$ and $S_x(E_y)$ induced spin components demonstrate the well-known transverse
in-plane character of the induced spin for linear Rashba SO coupling. It should be
noted that the local probe measurements of induced spin polarizations (or magnetization)
may detect the non-zero static and dynamic local magnetization \cite{Neudert2005} in the spot under
the probe even in case of total zero mean spin value (\ref{sn}). The examples of such systems with zero total
spin polarization but non-zero spin spatial distribution (spin textures) can be found
among the models of semiconductor superlattices with SO coupling both with\cite{PSK}
and without \cite{sooptic,spinelec} external magnetic field, but their experimental observation and device
application are currently limited by the probe and manipulation technology of the size of artificial
superlattices and quantum wells rather then probing and utilizing the spatial magnetic configurations
on the scale of individual atoms in the lattice. As to the predicted non-zero mean values of the induced
spin such as those predicted here for Bi/Si, they are related to the whole sample and
thus should be detectable. We believe that the predictions of the charge current and spin polarization generation
made in this Sec. can be useful in designing novel spintronic devices where the induced spin components are coupled
in a well-defined manner to the direction of the applied electric field, and this effect survives at room
temperature and finite collision rate.

\section{Spin polarization excited by electromagnetic field}

The response of the spin system in materials with significant SO coupling on the application
of an external electromagnetic radiation is among the most important and straightforwardly obtained
characteristics since the optical manipulation of spins is one of the main goals of spintronics
in general, and the linear response theory for the electromagnetic radiation effects is well-established
and easily applied. It was found in various papers that the spins with different
polarizations can be excited, depending on the symmetry of the electron Hamiltonian,
the type and strength of the SO terms, and on the polarization of the incident
radiation.\cite{Awschalom,Zutic,Bhat2005,Sherman2005,Tarasenko2005,Ganichev2007,sooptic,PSK}
As in the previous Sec., we shall calculate the response functions for the room
temperature and for a realistic collision broadening since the relatively large scale
of energy bands and SO splitting in the Bi/Si electron system compared to the
conventional GaAs, InGaAs or pure Si semiconductor structures can make Bi/Si being a promising
candidate for the observation and control of the predicted radiation-induced effects
in the devices operating even at room temperatures, as we hope.

The electromagnetic wave is considered to be propagating normally to the Bi/Si(111)
interface along the $z$ axis, and is characterized by the polarization of the electric
field vector ${\bf E}={\bf E}_0 \exp (i({\bf k} \cdot {\bf r}-\omega t))$
in the $(xy)$ plane, ${\bf E}_0=(E_{0x},E_{0y},0)$. In the dipole approximation
the interaction of the electromagnetic field with the electrons is described via the velocity
operators (\ref{vx}),(\ref{vx}) which include the SO part.
We start with the calculation of the absorption coefficient

\begin{eqnarray}
\alpha(\omega)=\frac{4\pi^2 e^2}{m^2 \omega c \sqrt{\varepsilon} V}
\sum_{n,n',{\bf k}} \mid \left({\bf e} \cdot {\bf v} \right)_{nn'{\bf k}} \mid ^2 \times \\
\times \delta (E_{n'{\bf k}}-E_{n{\bf k}}-\hbar \omega) \left(f_{n{\bf k}} - f_{n'{\bf k}} \right)
\label{absorp}
\end{eqnarray}

where ${\bf e}=(e_x,e_y,0)$ is the polarization vector for the incident wave, ${\bf v}=(v_x,v_y,0)$
is given by the velocity operators (\ref{vx}),(\ref{vy}), $f_{n{\bf k}}$ and $f_{n'{\bf k}}$ are Fermi equilibrium distribution
functions, $V$ is the sample volume, and the summation is taken over all energy bands $n$, $n'$ and the SBZ points
${\bf k}$. As we have already said, we don't know yet the exact values for many of the material parameters for Bi/Si
including the dielectric constant $\varepsilon$, and thus we shall focus mainly on the dependence of (\ref{absorp})
on the incident wave frequency and will scale the absolute value of absorption in arbitrary units which can always be
rescaled when the values material parameters will be clarified in future experiments.

The second quantity which frequency dependence we shall present together with the absorption is the induced
spin polarization $S_m(\omega)$ which can be derived by applying the Kubo linear
response theory\cite{Kubo1957,Callaway1974,PSK} and has the following form:

\begin{eqnarray}
S_m(\omega)=\frac{-i e E_{0l}}{8\pi m \hbar} \sum_{n,n',{\bf k}}
\frac{f_{n{\bf k}}-f_{n'{\bf k}}}{\omega_{nn'}({\bf k})} \times \\
\times \frac{S^{(m)}_{n'n} ({\bf k}) v^{(l)}_{nn'}({\bf k})}{\omega-\omega_{nn'}({\bf k})+i\nu}.
\label{sind}
\end{eqnarray}

Here the interband matrix elements of the spin $m$-th component operator
$S^{(m)}_{n'n}({\bf k})=\langle \psi_{{n' \bf k}} \mid {\hat \sigma}_m \mid  \psi_{{n \bf k}} \rangle $ as well
as the matrix elements for the $l$-th component of the velocity operator (\ref{vx}),(\ref{vy}) enter depending on the incident
wave polarization and on the desirable output for the spin component,
$\hbar \omega_{nn'}({\bf k})=E_{n{\bf k}}-E_{n'{\bf k}}$, the parameter $\nu$ is level broadening which we take as being
equal to the collision rate introduced in the previous Sec, and $V_c$ is the unit cell volume. As before, we shall
assume that $T=293$ K and $\nu=10^{12}$ $s^{-1}$ which should provide us with realistic expectations regarding
the absorption and induced spin dependence on the incoming photon energy.

\begin{figure}
  \centering
  \includegraphics[width=80mm]{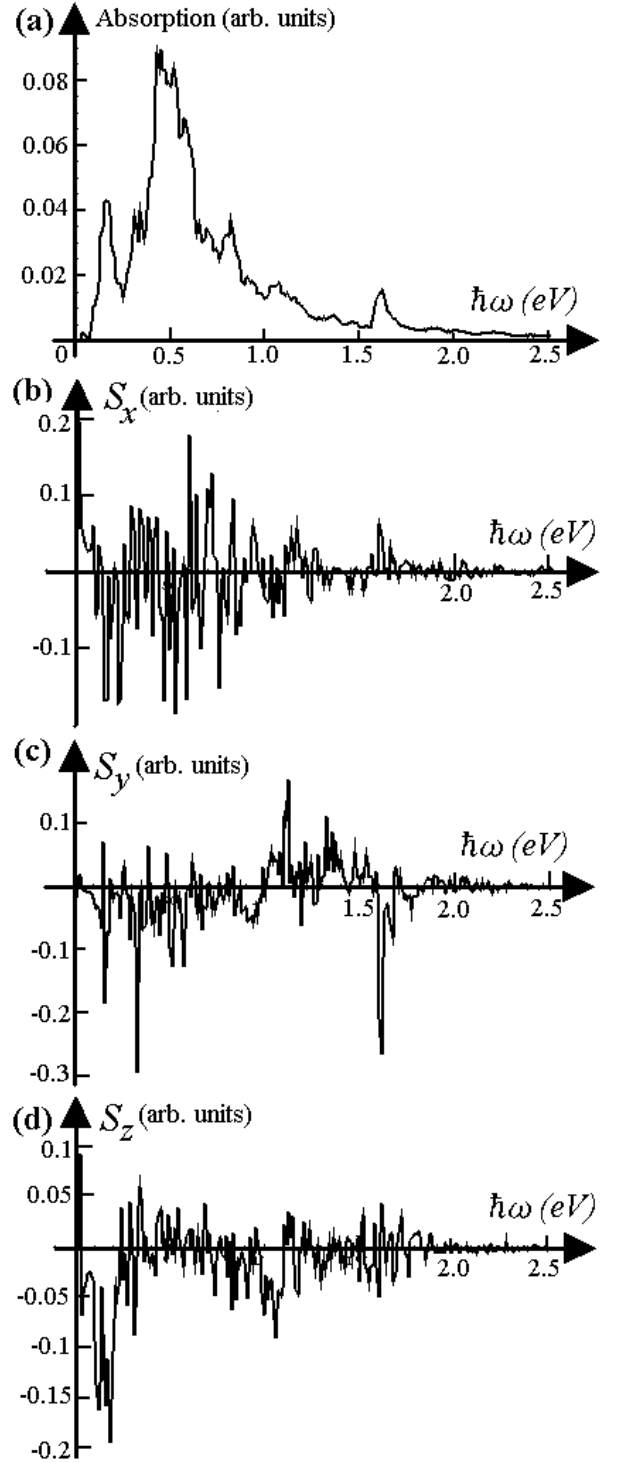}
  \caption{(a) Absorption coefficient and (b) - (d) components of the induced spin density in the Bi/Si surface
           electron gas  shown as a function of the incoming photon energy for the linear $x$-polarized incident
           radiation propagating normally to the interface plane. The greatest absorption and most of the spin
           polarization peaks are induced in the photon energy near $\hbar \omega \sim 0.5$ eV corresponding to
           the transitions between the highest occupied band 2 and the lowest unoccupied band 3 of the electron
           energy spectrum shown in Fog.\ref{fbands}. Both the in-plane spin components $S_x$, $S_y$ and
           the out-of plane component $S_z$ can be excited.}
  \label{fasx}
\end{figure}

\begin{figure}
  \centering
  \includegraphics[width=80mm]{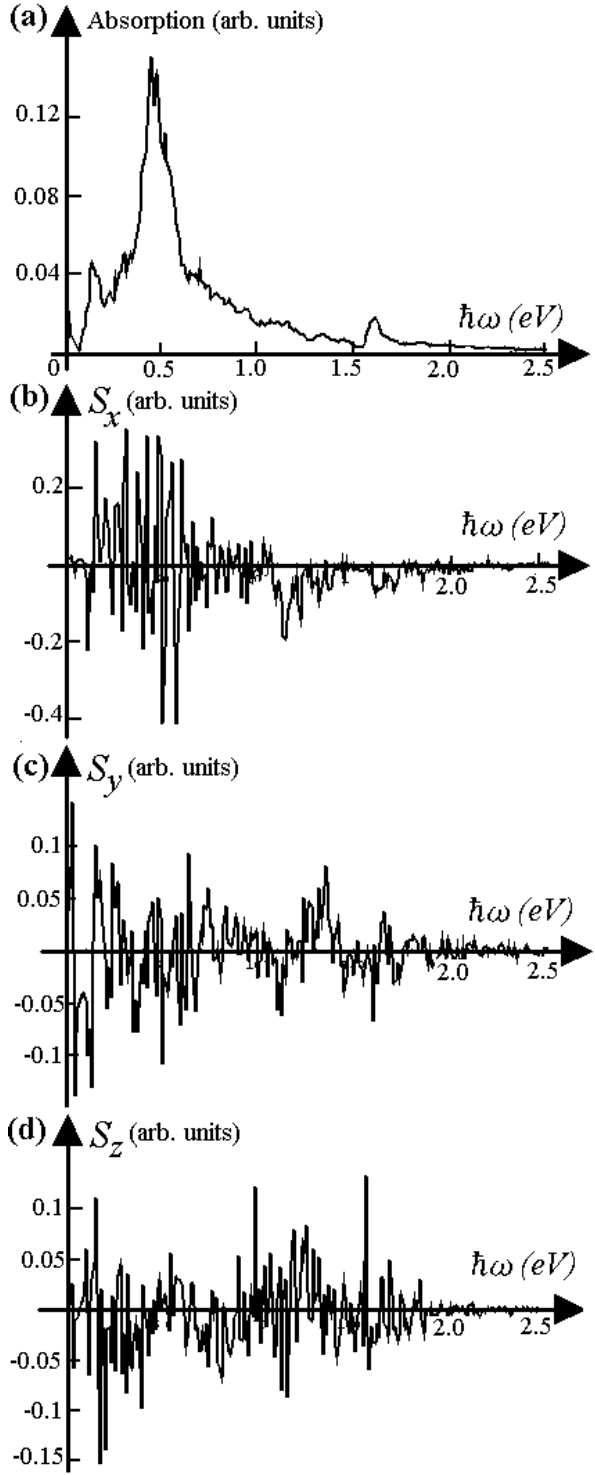}
  \caption{Photon energy dependence of (a) absorption coefficient and (b) - (d) components of the induced
           spin density for the same parameters as in Fig.\ref{fasx} but for the $y$-polarized incident radiation.
           The $x \leftrightarrow y$ lattice and energy band asymmetry in the hexagonal geometry is reflected in
           different shape and amplitude for the absorption coefficients in this Fig. and in Fig.\ref{fasx}.
           The Rashba term in SO coupling is reflected in the dominating excited $S_x$ component in this Fig. and
           the dominating $S_y$ component in Fig.\ref{fasx}.}
  \label{fasy}
\end{figure}

\begin{figure}
  \centering
  \includegraphics[width=80mm]{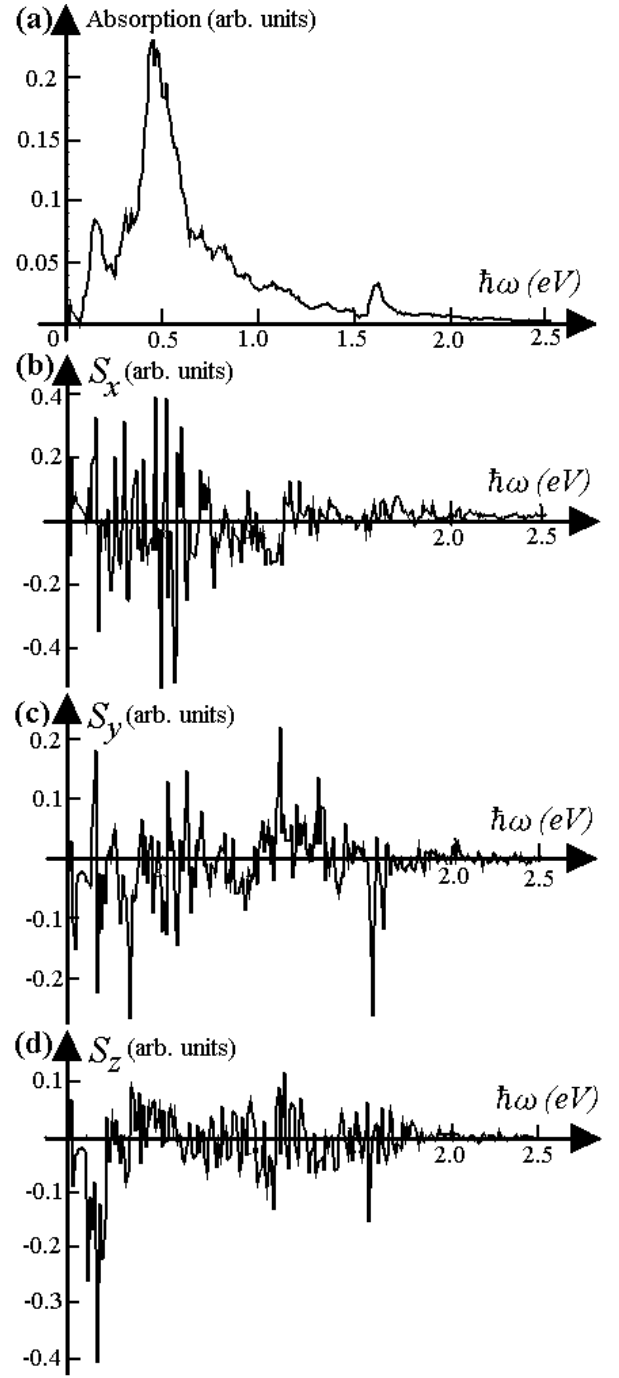}
  \caption{Photon energy dependence of (a) absorption coefficient and (b) - (d) components of the induced
           spin density for the same parameters as in Fig.\ref{fasx} but for the circular $\sigma^{+}$-polarized
           incident radiation. The common features of the response to both $x$-polarized and $y$-polarized
           incoming wave from Fig.\ref{fasx} and Fig.\ref{fasy} can be seen on the spin component figures.}
  \label{faspl}
\end{figure}

\begin{figure}
  \centering
  \includegraphics[width=80mm]{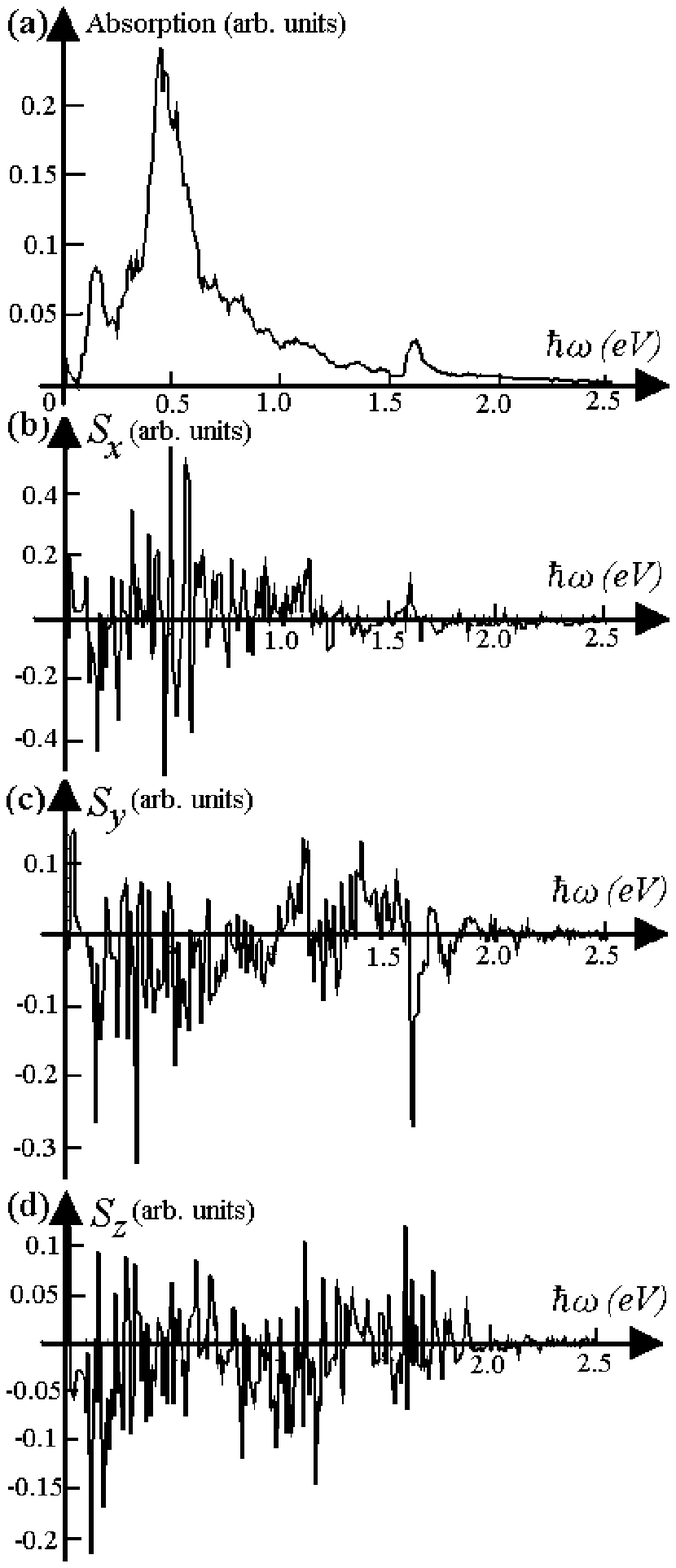}
  \caption{Photon energy dependence of (a) absorption coefficient and (b) - (d) components of the induced
           spin density for the same parameters as in Fig.\ref{fasx} but for the circular $\sigma^{-}$-polarized
           incident radiation. The absorption coefficient is unchanged from the $\sigma^{+}$ case in Fig.\ref{faspl}(a)
           while the excited spin components demonstrate quantitative differences while maintaining the same
           type of shape for the photon energy dependence.}
  \label{fasmn}
\end{figure}

The results for the photon energy dependence of the absorption coefficient (\ref{absorp})
and the induced spin polarization (\ref{sind}) are shown in Fig.\ref{fasx} through Fig.\ref{fasmn}
for the incident wave linearly polarized along $x$ (Fig.\ref{fasx}), along $y$
(Fig.\ref{fasy}) directions, and for circular $\sigma^{+}$ (Fig.\ref{faspl}) and
$\sigma^{+}$ (Fig.\ref{fasmn}) polarizations. The photon energy interval is chosen to cover
the whole energy band range of the four lowest bands shown in Fig.\ref{fbands} where
the most effective transitions occur between the states below and above the Fermi level.
One can see that both the in-plane spin components $S_x$, $S_y$ and the out-of plane component $S_z$ can be excited
on a comparable scale which is a distinguishable feature of the lower hexagonal symmetry combined with Rashba
SO coupling compared with one-dimensional\cite{Kleinert2005,sooptic,spinelec} or two-dimensional square\cite{PSK}
lattices with Rashba SO coupling.

If one compares the results for $x$- and $y$-polarized
radiation in Fig.\ref{fasx} and Fig.\ref{fasy}, it can be seen that, similar to the dc current properties discussed
in the previous Sec., the $x \leftrightarrow y$ lattice and energy band asymmetry in the hexagonal geometry of
the whole problem is reflected here in different shape and amplitude for the absorption coefficients
in Fig.\ref{fasx}(a) and \ref{fasy}(a). Again, by looking onto the relative magnitude of the excited spin components
in Fig.\ref{fasx}(b)-(d) and in Fig.\ref{fasy}(b)-(d) one can see that the pure Rashba SO coupling is reflected
in the dominating excited $S_x$ component in Fig.\ref{fasx} and correspondingly in the dominating $S_y$ component
in Fig.\ref{fasx}, i.e., in the in-plane and transverse direction relative to the electric field vector of the incident
wave. As to the results for the circular $\sigma^{\pm}$-polarized radiation presented
in  Fig.\ref{faspl} and Fig.\ref{fasmn}, on can see that the common features of the response to both $x$-polarized and
$y$-polarized incoming wave from Fig.\ref{fasx} and Fig.\ref{fasy} can be seen on the spin component figures since both
$v_x$ and $v_y$ operators here enter the expressions (\ref{absorp}) and (\ref{sind}) for the response.
The absorption coefficient is totally insensible to the direction of rotation for the
incoming wave as it can be seen by comparing Fig.\ref{faspl}(a) and Fig.\ref{fasmn}(a).
The shape of the photon energy dependence for the excited spin components in Fig.\ref{faspl}(b)-(d) and
Fig.\ref{fasmn}(b)-(d) is different for $\sigma^{+}$ and $\sigma^{-}$ polarizations,
but these differences have a quantitative rather than a qualitative character since
the hexagonal symmetry of the lattice and the energy bands does not make any of these
two polarizations preferable from the point of view of the response quantities (\ref{absorp}) and (\ref{sind}).

In conclusion, the calculation and analysis of the absorption and spin polarization
response to the monochromatic electromagnetic radiation with normal incidence and
having different polarizations demonstrates that this radiation is most effectively
absorbed in the photon energy range of around $0.5$ eV corresponding to the photon wavelength
$\lambda =2.47$ $\mu m$ where both in-plane and out-of-plane spin components can be excited
at realistic temperature and collision broadening on a comparable scale with relative amplitudes depending
on the precise value of frequency and the polarization of the incident radiation.
These properties can be useful for designing new optical and spintronic devices coupling
the electron spin with light and operating at room temperature.

\section{Conclusions}

We have developed an expansion of the nearly free-electron model\cite{FPG} describing the energy bands
and spin polarization for the electron states at Bi/Si(111) interface with giant spin-orbit coupling,
and applied it for the linear response analysis for charge current and induced spin caused by dc field
and by electromagnetic radiation. It was found that the large spin-orbit coupling in this system may allow
resolving the spin-dependent properties even at room temperature and at realistic collision rate.
The geometry of the atomic lattice combined with spin-orbit coupling leads to an anisotropic response both
for current and spin components related to the orientation of the external field.
The in-plane dc electric field produces only the in-plane components of spin in the sample while both
the in-plane and out-of-plane spin components can be excited by normally propagating electromagnetic
wave with different polarizations. The qualitative predictions of the charge and spin
response in a novel and promising Bi/Si system may be useful for the forthcoming
detailed theoretical and experimental studies which may lead to the development of
principally new electronic, optical and spintronic devices operating at room
temperature. Further theoretical and especially experimental studies of this promising
system with big SO coupling allowing the survivability of the spin-related effects at
room temperature are expected bringing us new and fascinating phenomena with both
fundamental, experimental and device-related results.

\section*{Acknowledgments}

The author is grateful to V.Ya. Demikhovskii, A.M. Satanin, A.A. Perov for helpful discussions,
and to A.A. Chubanov for technical assistance. The work was supported by the RFBR Grants
11-02-00960a and 11-02-97039/Regional, and by the RNP Program of Ministry of Education and Science RF.

\end{document}